\documentclass[pra,groupedaddress,twocolumn]{revtex4}
\usepackage{amssymb}
\usepackage{graphicx}
\usepackage{dcolumn}
\usepackage{bm}
\usepackage{amsmath}
\usepackage{subfigure}
\usepackage{float}
\usepackage{color}
\usepackage{epstopdf}
\usepackage{upgreek}
\usepackage{hyperref}
\newcommand{\EF}{\mathcal{E}}

\begin{document}

\title{Optical control of atom-ion collisions using a Rydberg state}

\author{Limei Wang$^1$}
\author{Markus Dei{\ss}$^1$}
\author{Georg Raithel$^{2}$}
\author{Johannes Hecker Denschlag$^1$}
\thanks{Corresponding author: johannes.denschlag@uni-ulm.de}
\affiliation{$^1$Institut f\"ur Quantenmaterie and Center for Integrated Quantum Science and Technology IQ$^\mathrm {ST}$, Universit\"at Ulm, 89069 Ulm, Germany\\
$^{2}$Department of Physics, University of Michigan, Ann Arbor, Michigan 48109-1120, USA}
\date{\today}

\begin{abstract}
We present a method to control collisions between ultracold neutral atoms in the electronic ground state and trapped ions. During the collision, the neutral atom is resonantly excited by a laser to a low-field-seeking Rydberg state, which is repelled by the ion. As the atom is reflected from the ion, it is de-excited back into its electronic ground level. The efficiency of shielding is analyzed as a function of laser frequency and power, initial atom-ion collision energy, and collision angle. The suitability of several Rydberg levels of Na and Rb for shielding is discussed. Useful applications of shielding include the suppression of unwanted chemical reactions between atoms and ions, a prerequisite for controlled atom-ion interactions.
\end{abstract}

\maketitle
\section{Introduction}
The developing field of hybrid systems of cold, trapped atoms and ions (for reviews see, e.g., \cite{Haerter2014a, Willitsch2015, Tomza2017}) has brought forth a number of proposals for novel experiments. For many of these, including proposals described in \cite{Casteels2011,Cote2002,Bissbort2013,Joger2014, Doerk2010}, any reactions between atoms and ions are unwanted and need to be suppressed.

One way to prevent atoms and ions from reacting with each other is to keep them at a sufficiently large distance by tightly confining them in individual traps \cite{Idziaszek2007}. Another approach is optical shielding, a method that has been introduced about twenty years ago as a tool to reduce inelastic losses in samples of ultracold neutral atoms \cite{Band1995,Suominen1995,Napolitano1997,Katori1994,Marcassa1994,Sanchez1995,Walhout1995}. There, a blue-detuned laser induces a transition of a colliding atom pair to an electronically excited repulsive molecular potential. Collisional suppression of up to a factor of 30 has been demonstrated \cite{Walhout1995}. Recently, the emergence of cold molecules has stimulated a renewed interest in optical shielding \cite{Gorshkov2008,Zhao2012} and reaction control, see e.g. \cite{Mills2019}. These proposed schemes for shielding rely on Rydberg-dressing, i.e. laser-admixing of Rydberg levels to the ground state. In this approach, large dipolar interactions between Rydberg atoms are utilized to generate strong repulsion. Rydberg-dressing was also recently proposed for suppressing collisions between ultracold neutral atoms and ions \cite{Secker2017}, where the Rydberg dressing operates on a forbidden $S$ to $S$ transition. As the particles approach each other, the electric field of the ion increasingly admixes $P$-character into the Rydberg $S$-level, leading to increasing optical coupling between the Rydberg level and the $S$ ground level. This generates a repulsive ac-Stark-shift potential barrier at small internuclear separations.

In the present work, we propose a distinct optical shielding scheme for atom-ion collisions. Our method is based on an adiabatic optical transition of the atom from its ground state towards a low-field-seeking Rydberg state, as the atom traverses the ion's electric field. We analyze the efficiency of the shielding process as a function of laser frequency, laser power, the initial collision energy and the collision angle. Our proposed scheme offers particularly interesting opportunities when the atomic ground state is optically coupled into a manifold of Rydberg Stark states which contains avoided crossings. In this case, the collision dynamics occurs on coupled potential landscapes, with mixed adiabatic and non-adiabatic evolution as well as tunneling playing an important role.

\section{Generic shielding scheme }
\label{sec:simplescheme}

\begin{figure}[t]
	\centering
	\includegraphics[width=0.4\textwidth] {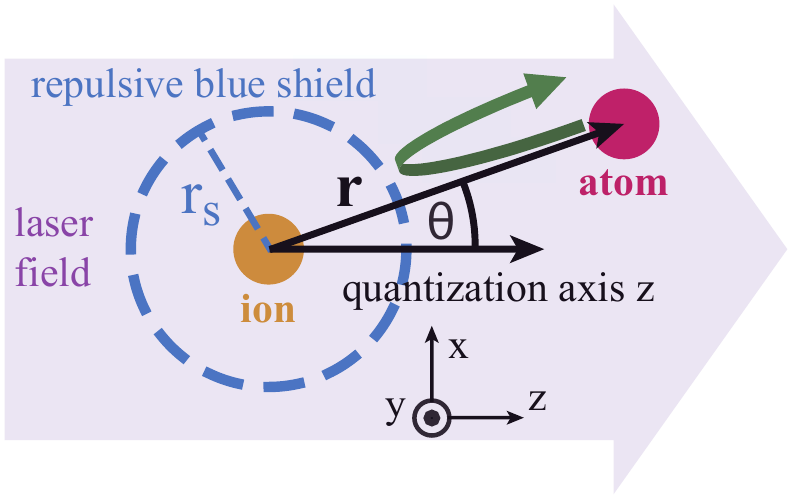}
	\centering
	\caption{Blue shielding scheme using a Rydberg level. In a collision a neutral ground state atom approaches an ion. At a distance $r \approx r_s$ a cw laser adiabatically excites the neutral atom to a Rydberg level. The Rydberg atom is in a low-field-seeking state and is strongly repelled by the electric field of the ion. Effectively, the atom is reflected from a potential barrier around the ion. As the reflected atom leaves, it is adiabatically de-excited back into the ground state.}
	\label{Fig.1}
\end{figure}

We consider an atom in the electronic ground state  $|g \rangle$ which collides with a trapped ion in a two-body collision, see Fig.$\:$\ref{Fig.1}. The atom-ion pair is located in an intense continuous-wave (cw) laser field. The distance between atom and ion is denoted $r$. When $r$ reaches the shielding distance $r_s$ the atom is resonantly excited to a Rydberg state $|e \rangle$ by the  laser field. The Rydberg state has a large low-field-seeking electric dipole moment which leads to a repulsion of the collision partners, such that the atom is effectively reflected off a potential wall at distance $r_s$. After the reflection the atom is adiabatically de-excited back to the ground state. The collision takes place on a time scale that is much shorter than the natural lifetime of the Rydberg state of several $\upmu$s. Therefore, spontaneous radiative decay of the Rydberg atom is negligible.

\begin{figure}[htb]
	\centering
	\includegraphics[width=0.45\textwidth] {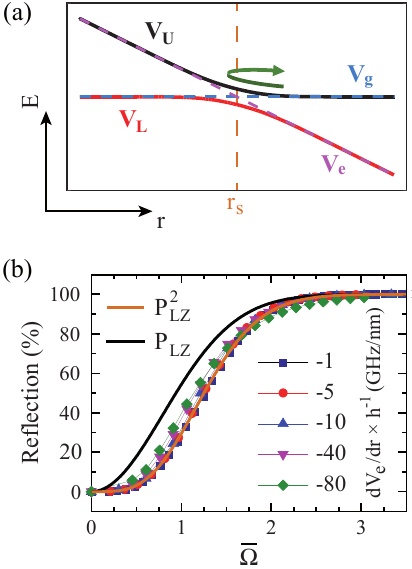}
	\centering
	\caption{(a) Illustration of the blue shielding scheme exploiting an avoided crossing.
		Shown are the potential energies in a photon-dressed picture of the neutral atom at a distance $r$ from the ion. $V_g$ (blue dashed line) denotes the potential energy curve for the atom in the ground state. $V_e$ (dashed purple curve) is the curve for a low-field-seeking, repelled Rydberg state. The coupling of the two levels by the laser leads to an avoided crossing at $r_s$. The adiabatic potential energy curves are $V_U$ (black solid line) and $V_L$ (red solid line). (b) Reflection probability as a function of the scaled coupling strength $\bar{\Omega}$ for a collision energy $E_{coll}=1\:\text{mK}\times k_\text{B}$ and slopes $dV_e/dr$ given in the legend. The black (orange) solid line
corresponds to the probability $P_{LZ}$ ($P_{LZ}^2$) for one (two) adiabatic passages through the crossing according to the Landau-Zener formula, respectively (see text).  }
	\label{Fig.3}
\end{figure}

We estimate the shielding efficiency of this scheme using a two-channel model with ground state $|g \rangle$ and excited Rydberg state $|e \rangle$. Using the rotating wave approximation the interaction Hamiltonian in the rotating frame is
\begin{equation}
V_I (r)=
\left( \begin{array}{cc}
V_{g}  &  \hbar\Omega /2 \\
\hbar\Omega /2	\, \, \, \,   &   (r-r_s) \, dV_e/dr
\end{array}\right)\,,
\label{eq:VIlinear}
\end{equation}
where the ground-state channel $|g \rangle$ has a constant potential energy $V_g = 0$ (the interaction between the ground-state atom and the ion is neglected). We assume for now that the potential energy of the low-field-seeking Rydberg channel $|e \rangle$ has a constant slope $dV_e /dr$, as depicted in Fig.$\:$\ref{Fig.3}(a). The optical coupling $\Omega$ of the two channels leads to an avoided crossing around $r_s$. For a sufficiently small atom-ion collision energy $E_{coll}$, the atom adiabatically enters the avoided crossing on its way in, climbs the potential slope of the Rydberg state $|e \rangle$, turns around, and adiabatically leaves the avoided crossing on its way out. Quantitatively, we solve the radial Schr\"{o}dinger equation
\begin{equation}
\left(\! -\frac{\hbar^2}{2\mu} \frac{d^2}{dr^2}  + V_I(r) \right)  \Psi(r)  = E_{coll}  \Psi(r) =  E_{coll}
\left(\!
\begin{array}{c}
\varphi_{g}(r) \\
\varphi_{e}(r) \\
\end{array}\!
\right)\!\,.
\label{eq:SE}
\end{equation}
Here, $\mu$ is the reduced mass of atom and ion, and $\Psi(r)$ is the two-component wave function of the system. For simplicity, we assume that the atom-ion collision takes place in an $s$-partial wave, and that the laser coupling between the atomic ground and Rydberg state is isotropic. We calculate scattering solutions for this one-dimensional problem with an incident probability flux in the $|g \rangle$-channel. The Schr\"{o}dinger equation is numerically integrated for a chosen collision energy $E_{coll}$, as described in sections \hyperref[sec:NumSolution]{A} and \hyperref[sec:AdiabaticBasis]{B} of the Appendix.
From the solution $\Psi(r)$ we find the incoming and outgoing fluxes by first expressing the wave function components as $\varphi_{e,g}= a e^{ikr}+b e^{-ikr}$ with $k = \sqrt{2 \mu E_{coll}}/ \hbar$, at a location $r \gg r_s$. For example, for the ground state, the coefficients are $a=(d \varphi_{g} / d r+ ik \ \varphi_{g}) /(2ik e^{ikr})$ and $b=(d \varphi_{g}/ d r - ik \ \varphi_{g}) /(-2ik e^{-ikr})$. The reflection probability of the incident ground-state atoms from the blue shielding potential is $|a|^2/|b|^2$. This is plotted in Fig.$\:$\ref{Fig.3}(b) as a function of a scaled coupling strength
\begin{equation}
\bar{\Omega} = \hbar \Omega \, \,  \left( { 2 \mu \over \hbar^2 E_{coll} } \right)^{1/4} \, \, |dV_e/dr|^{-1/2}
\label{eq:OmegaBar}
\end{equation}
for a collision energy $E_{coll} = 1\:\text{mK} \times k_\text{B}$ and several slopes $dV_e/ dr$ (see legend). We note that in a range of reflection probabilities between about 10 and 90\% all curves for the different slopes $dV_e/ dr$ are quite similar. They can therefore approximately be described by a universal function.
Indeed, the squared Landau-Zener probability for adiabatic transfer
$P_{LZ}^2=(1- \exp{(-\pi \bar{\Omega}^2/4)})^2$ describes the data quite well \cite{Landau1932, Zener1932}. This goes along with two adiabatic passages, forth and back, as indicated by the green arrow in Fig.$\:$\ref{Fig.3}(a).

\begin{figure}[htb]
	\centering
	\includegraphics[width=0.45\textwidth] {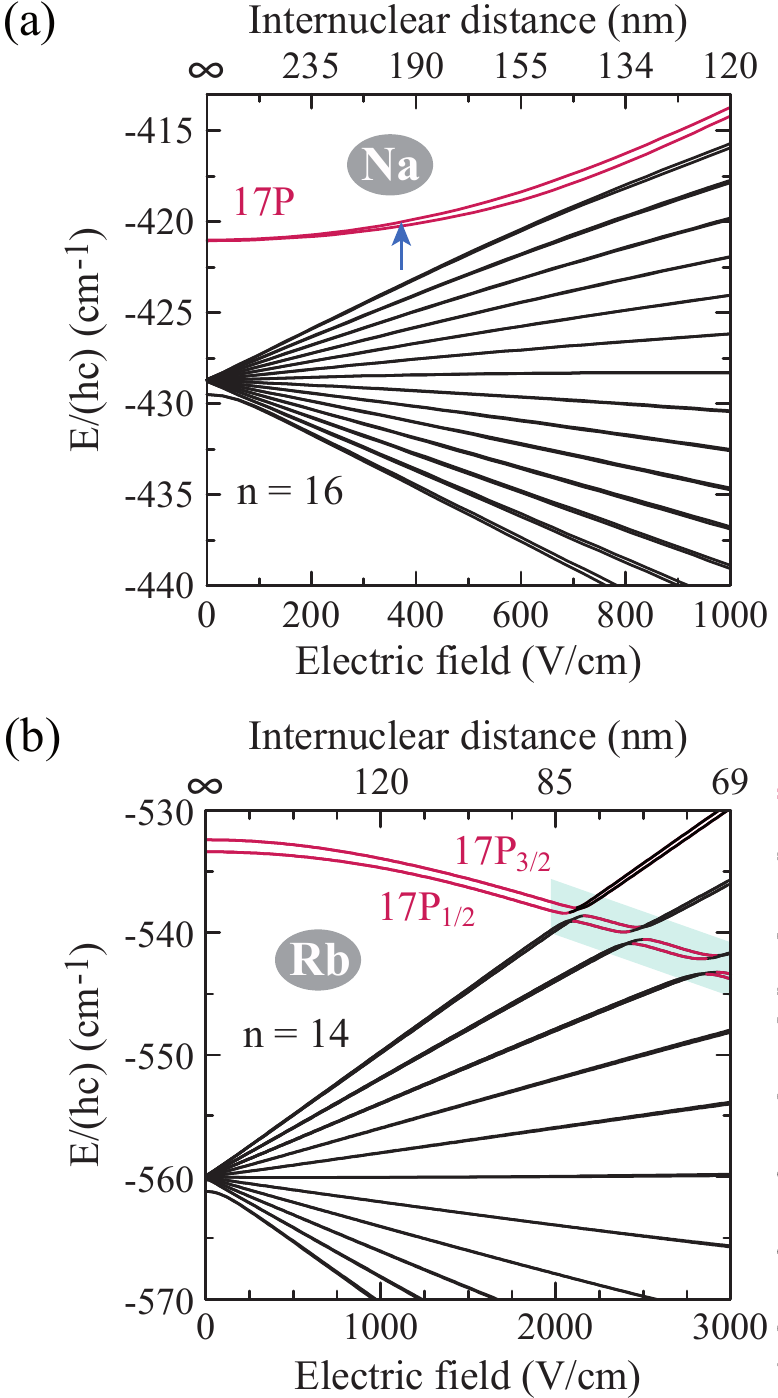}
	\centering
	\caption{(a) Stark spectrum of Na in the vicinity of the low-field-seeking state $17P$.	(b) Stark spectrum of Rb in the vicinity of the high-field-seeking states $17P_J$, which have a (zero-field) fine-structure splitting of about 1$\:$cm$^{-1}$. Avoided crossings of the $17P_J$-states with low-field-seeking states of the $n=14$-manifold occur within the cyan-shaded area.}
	\label{Fig.2}
\end{figure}

A possible candidate for Rydberg shielding as described is Na. Figure~\ref{Fig.2}(a) shows a Stark map of the $n$ = 16 hydrogen-like manifold (i.e. levels with orbital angular momentum quantum number $L > 1$) as well as the 17$P$-level which is a low-field seeker. Here, $n$ denotes the principal quantum number. The 17$P$-level has a natural lifetime of about 5$\:\upmu$s, including decay due to black body radiation \cite{Saffman2005}. By choosing a laser frequency we set both the ion electric field at which the transition to 17$P$ is resonantly driven as well as the slope $dV_e/dr$. The electric field of a singly charged ion is related to the internuclear distance via $E_f = e /(4\pi \varepsilon_0 r^2)$ \cite{Note1}. Here, $e$ is the elementary electric charge, and $\varepsilon_0$ is the dielectric constant of the vacuum. As an example for possible shielding parameters for Na we consider laser excitation to an energy of $-420\:$cm$^{-1}$ in Fig.$\:$\ref{Fig.2}(a), which is blue-detuned by 1~cm$^{-1}$ from the zero-field location of the 17$P$ state. In that case, the shielding radius is $r_s \approx 200\:\text{nm}$ [blue arrow in Fig.$\:$\ref{Fig.2}(a)] and $dV_e/dr \approx - h \times 0.6\:$GHz/nm. For a Rabi frequency $\Omega = 2\pi \times 400\:$MHz the scaled coupling strength is $ \bar \Omega \approx 2 $, and the reflection probability is $> 90\%$ [see Fig.$\:$\ref{Fig.3}(b)].

The situation is richer and more interesting in atomic species with high-field-seeking $P$ states, such as Rb. A Stark map for Rb 17$P$ is shown in Fig.$\:$\ref{Fig.2}(b). While high-field-seeking states are normally not suitable for Rydberg shielding with our scheme, even in that case shielding can still be achieved by utilizing avoided crossings between the $17P_J$-states and the $n=14$ hydrogen-like manifold. Figure \ref{Fig.2}(b) shows several such avoided crossings. Close to the crossings, the low-field-seeking $n=14$ hydrogen-like states exhibit substantial $P$-character, making them excitable from the atomic ground state $|g  \rangle$ and enabling efficient shielding schemes.

In the following we will study in detail how Rydberg shielding works right at the location of these avoided crossings. In Fig.$\:$\ref{Fig.2}(b) the first avoided crossings occur at an electric field of about 2100$\:$V/cm. This corresponds to a reflection distance $r_s$ of about 80$\:$nm. In order to resonantly couple to these avoided crossings, a laser wavelength of about $301.7\:\textrm{nm}$  is needed. In principle, $r_s$ can be tuned over a large range by choosing an appropriate avoided crossing in another manifold with a different $n$-quantum number. We find that
$r_s  \approx 0.079\:\text{nm} \times  n_{\text{eff}}^{2.59}$, where $n_{\text{eff}}= n-\delta(n)$ is the effective principal quantum number and $\delta(n)$ is the quantum defect \cite{Note2}. For $n$-quantum numbers between 13 and 26, the shielding radius $r_s$ ranges from 33 to 269$\:$nm.

\section{A case study}
 In the previous section we have briefly sketched the physics behind Rydberg shielding, which has involved a number of simplifications and assumptions. We now proceed with a more realistic case study, which requires more detail.

\subsection{Ground state polarization potential, centrifugal barrier and collision energy}
\label{sec:CaseStudyA}

In Eq.$\:$(\ref{eq:VIlinear}) we have neglected the interaction potential $V_g$ between ground state atom and ion. This is indeed justified for our purpose as we show in the following.  The long-range tail of $V_g$ (see e.g. Ref.$\:$\cite{Haerter2014a}) is given by
\begin{equation}
V_g(r, l)=- \frac{C_4}{2r^4}+\frac{\hbar^2l(l+1)}{2\mu r^2}\,,
\label{gspot}
\end{equation}
where the first term represents the polarization potential of the ground state $|g \rangle$ in the ion electric field. The second term is the centrifugal potential for the internuclear motion of the ion-atom system with $l$ denoting the quantum number of the partial wave. For the $^{87}$Rb $5S_{1/2}$ state, $C_4= \alpha e^2/(4\pi\varepsilon_0)^2$ with the static dipolar polarizability $ \alpha= 4\pi \varepsilon_0  \times 4.74\times10^{-29}\:$m$^3$ \cite{Gregoire2015}.

In current experiments, the collision energy of a cold ion in a Paul trap colliding with an ultracold atom is typically on the order of 1$\:\text{mK} \times k_{\text{B}}$, due to effects of micromotion of the ion. This is indeed much larger than the  $\approx 10\:\upmu$K$\times k_{\text{B}}$  depth of the polarization potential at the shielding distance $r_s \approx 80\:$nm. Therefore, the polarization potential in the ground state can be safely neglected.

At the same shielding distance ($r_s \approx 80\:$nm) the centrifugal potential reaches a height of 1$\:\text{mK} \times k_{\text{B}}$ for $l = 38$. Therefore, a large number of partial waves are involved in a typical atom-ion shielding experiment. Nevertheless, if shielding works for the $s$-wave it will also work for the higher partial waves because the centrifugal potential only helps the shielding. Further, the thermal energy of 1$\:\text{mK} \times k_{\text{B}}$ is orders of magnitude lower than the variation of the Rydberg levels over the relevant range of $r$. Therefore, we can quite generally restrict ourselves to the discussion of only the $s$-wave. Doing so, we neglect mixing of partial waves, which occurs because the atom-ion interaction is in general not spherically symmetric. However, this   mixing  is not relevant for shielding and therefore beyond the scope of this work.

Finally, we would like to mention that besides resonantly coupling the ground state and the Rydberg state the shielding laser also produces an optical dipole potential for the neutral ground state atoms.
This dipole potential is repulsive and amounts to
$\sim2\:$mK$\times k_\text{B}$ \cite{Grimm2000} for the typically needed laser intensities in our shielding scheme.
 Since this is on the order of the collision energy, it needs to be considered in experimental work. In principle, the repulsive dipole potential can be compensated by applying an additional attractive dipole trap (e.g., based on a 1064$\:$nm laser).

\subsection{Interaction between a Rydberg atom and an ion}
\label{sec:AtomIonInteraction}
In section \ref{sec:simplescheme} the interaction between a Rydberg atom and an ion was approximated at several instances. We now refine the model by taking into account the inhomogeneity of the electric field of the ion. The field of the ion polarizes the Rydberg atom by mixing various orbital angular momentum states. This turns the Rydberg atom into an electric multipole which is either attracted or repelled by the ion.
For convenience we use in this section a coordinate system where the Rydberg atom is located at the origin.
The ion is located at $(0,0,-r)$ (in cartesian coordinates). Thus, the angle $\Theta$ in Fig.~\ref{Fig.1} is zero and the $z$-axis (quantization axis) is the internuclear axis.
The relative Rydberg-electron coordinates are denoted $(r_e, \theta_e, \phi_e)$.
Multipole expansion \cite{J.D.Jackson} of the electrostatic potential energy of the Rydberg electron and the positively charged point-like Rydberg atom nucleus within the field of the ion yields
\begin{equation}
V_{e+ion,ion} = -\frac{e^2}{4\pi\varepsilon_0}\sum_{l=1}^{\infty}\sqrt{\frac{4\pi}{2l+1} }  \frac{ r_e^l }{ {r}^{l+1} } Y_{l0}(\theta_e, \phi_e)\, .
\label{eq_ion-atom}
\end{equation}
Note that $Y_{l0} = Y^\ast_{l0}$.
\begin{figure}
	\centering
	\includegraphics[width=\columnwidth] {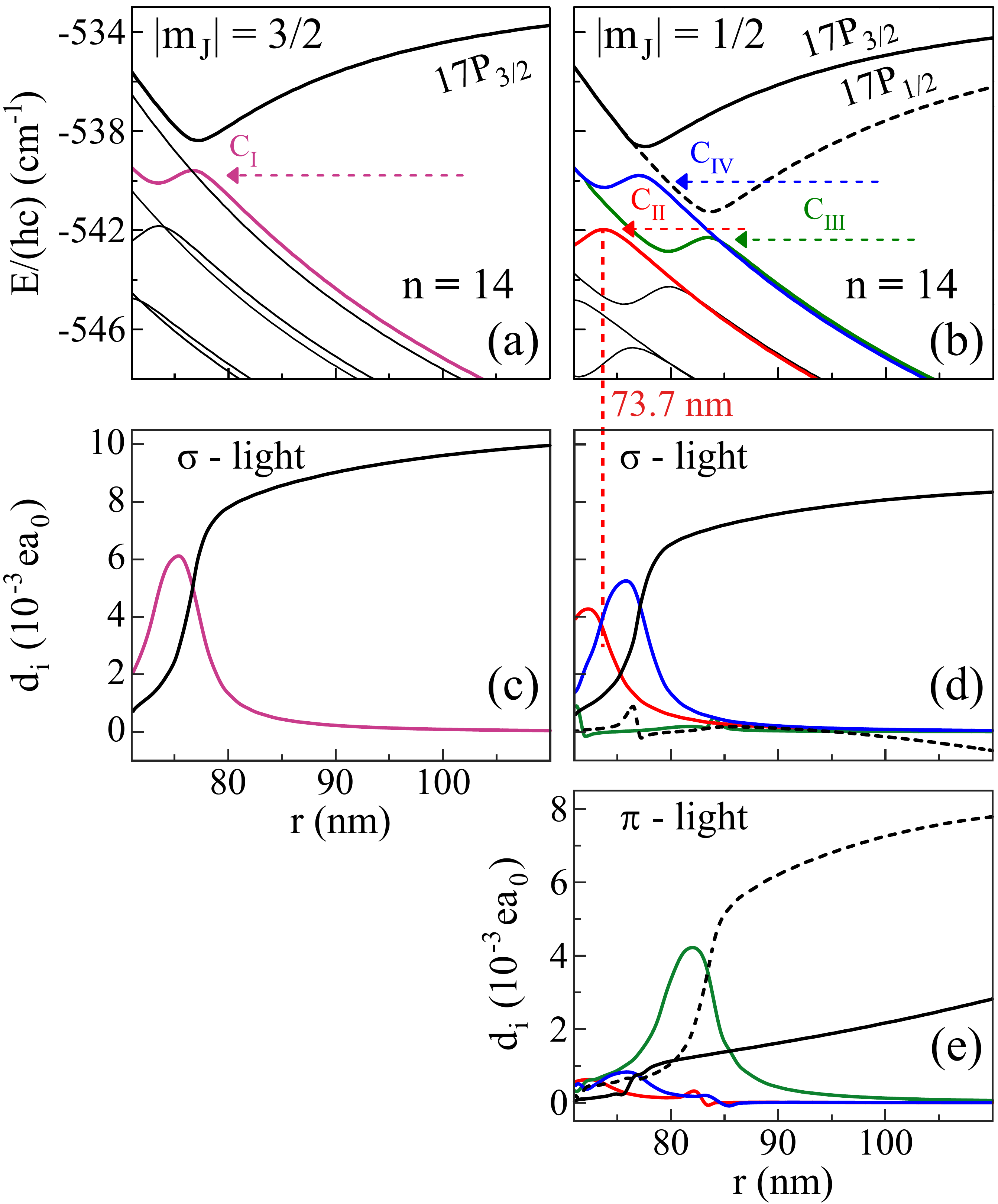}
	\caption{(a) and (b) Potential energy curves $V_{e,k}$ vs. internuclear separation $r$ for a structureless, singly charged ion interacting with a $^{87}$Rb Rydberg atom. The internuclear axis is the quantization axis. $m_J$ is the magnetic quantum number of the Rydberg atom. Shown are $17P_J$ states and levels of the $n=14$ hydrogen-like manifold. The level crossings $C_i$ are discussed in the text. (c), (d) and (e) Calculated transition electric dipole moments  $d_{i}(e_k)$ for transitions from the electronic ground state $|g \rangle$. The color coding for (c) is the same as in (a) and for (d), (e) it is the same as in (b).
	}
	\label{fig:PotCurvesDipoleMoments}
\end{figure}
The lowest-order term in Eq.~(\ref{eq_ion-atom}) corresponds to $l=1$ and treats the atom-ion interaction as if the ion produced a locally homogeneous electric field at the location of the atom.
This would give rise to a level pattern equivalent to that of the usual Stark effect, as in Fig.~\ref{Fig.2}, with the electric field at the Rydberg-atom center, $e/(4 \pi \varepsilon_0 r^2)$, plotted on the $x$-axis.
In our improved model, we include all terms in the sum of Eq.~(\ref{eq_ion-atom}) up to $l=6$. It is found that higher-order terms lead to negligible contributions. We obtain the potential energy curves (PECs) $V_{e,k}(r)$ for the Rydberg-atom-ion system by solving the Schr\"{o}dinger equation for the electron motion,
\begin{equation}
\left( \hat{H}_0 + \hat{V}_{e+ion,ion}(r) \right)  \vert e_k (r) \rangle   =  V_{e,k}(r)  \vert e_k (r) \rangle \,,
\label{eq_Schroedeq}
\end{equation}
 using a dense grid of internuclear separations $r$ that are held fixed in each calculation. Here, $\hat{H}_0$ is the  Hamiltonian of the unperturbed Rydberg atom including fine structure and $k$ is a label for the numerous PECs of the system.
 The electronic eigenstates $\vert e_k (r) \rangle$ have good magnetic quantum numbers $m_J$.
The total angular momentum $J$ becomes  good  at large enough atom-ion distances $r$.
In the framework of the Born-Oppenheimer approximation, the PECs $V_{e,k}(r)$ govern the radial (vibrational) dynamics of the Rydberg-atom-ion system.
Figures \ref{fig:PotCurvesDipoleMoments} (a) and (b) present numerical calculations of the $V_{e,k}(r)$ potentials for  $^{87}$Rb, again for the hydrogen-like $n=14$ manifold,  together with the fine-structure-split levels $17P_{3/2}$ and $17P_{1/2}$. The fine structure also causes the (much smaller) doublet structure within the hydrogenic $n=14$ manifold.
Shown are the results for magnetic quantum numbers $m_J=\pm 3/2$ (a) and $m_J=\pm 1/2$ (b) of the Rydberg atom.

\subsection{Transition electric dipole moments and Rabi frequency }
\label{sec:TransDipole}
Figures $\:$\ref{fig:PotCurvesDipoleMoments}(c), (d) and (e) show calculations of the transition
electric-dipole matrix elements
\begin{equation}
  d_{i}(e_k) =   \langle e_k (r), m_J | (-e) r_{e,i}|g, m\rangle
  \label{eq:diek}
\end{equation}
for transitions from the electronic ground state  $|g, m \rangle $ where $m = \pm 1/2$ is the magnetic quantum number of the
  angular momentum $J = 1/2$
\footnote{ Due to hyperfine interaction, the quantum state of a ground state $^{87}$Rb atom is normally described in terms of the quantum numbers $F, m_F$ corresponding to the total angular momentum  $\vec{F}=\vec{J}+\vec{I}$ with nuclear spin $\vec{I}$. Pure $m=\pm 1/2$ states can be prepared by working with the spin stretched states $| F = 2, m_F = \pm 2 \rangle$. }.
To easily distinguish the $m$-quantum numbers for excited and ground states, we do not attach an index $J$ to $m$ for the ground state.
 The $r_{e,i}$ are the spherical components of the electron coordinates (see section \hyperref[sec:AppArbitAngle]{C} of the Appendix for more details). The index $i \in \{+1, 0 , -1\}$  indicates whether we consider a $\sigma^+, \pi, \sigma^- $- transition, respectively. The matrix elements $d_{i}(e_k)$ are only nonzero for $m_J = i + m$.
  For now the quantization axis $z$  coincides with the atom-ion internuclear axis. The results in Fig.$\:$\ref{fig:PotCurvesDipoleMoments}(c)-(e)   are given in units of $ea_0$, where $a_0$ is the Bohr radius.
 The signs of the dipole matrix elements are fixed by ensuring that for every state $\vert e_k \rangle$ the amplitude of the $17P_{3/2}$ component is positive.
For the energetically degenerate states $ | e_k, \pm m_J \rangle $ and $|g, \pm m \rangle$ one finds
  \begin{equation}
   \langle e_k,   m_J |  r_{e,i}|g, m \rangle   =   \langle e_k, -m_J |  r_{e,-i}|g, -m \rangle\,,
  \end{equation}
  i.e., $  d_{i}(e_k) =   d_{-i}(e_k)$.
 As expected, the transition matrix elements $d_{i}(e_k)$ vary markedly as a function of $r$, particularly in the vicinity of  the avoided crossings. This is a consequence of the $r$-dependence of the mixing between the $17P_J$-states and the hydrogen-like Rydberg levels.
 The Rabi frequency $\Omega$ for the coupling  of the ground state   $|g \rangle = \sum_{m} \beta_{m} |g, m\rangle$  to the  Rydberg state $|e_k, m_J\rangle$
is given by
\begin{equation}
{\Omega}(m_J) = \sum_{i,m} \frac{\EF_{i} d_{i} (e_k) \beta_m} {\hbar} \, ,
\end{equation}
where $\EF_{i}$, $i \in \{+1, 0 , -1\}$, are the spherical components of the light field driving
$\sigma^+, \pi, \sigma^- $ transitions, respectively.

We now consider the internuclear axis $z'$ between ion and Rydberg atom to form an angle $\Theta$ with the lab
frame's quantization axis, $z$, which is defined by the propagation direction of the laser.
For simplicity, we assume that a rotation by $\Theta$ about the $y$-axis rotates the lab frame into the molecular frame $\{x',y',z'\}$ (see Fig.$\:$\ref{Fig.1} and Appendix \hyperref[sec:AppArbitAngle]{C}). In the molecular frame (primed frame)
the atomic ground state   $|g \rangle = \sum_{m} \beta_{m} |g, m\rangle$  becomes $|g \rangle = \sum_{m'} \beta'_{m'} |g, m'\rangle$.
 The coefficients in the molecular frame, $\beta'_{m'}$, are related to those in the lab frame, $\beta_{m}$, via  $\beta'_{m'} = \sum_m d^{(1/2)}_{m',m}(-\Theta) \, \beta_{m}$. Here,
  $d^{(1/2)}_{m',m}(-\Theta) $
  are the elements of Wigner's small d-matrix.
The spherical components of the electric field transformed from the lab frame into the molecular frame are denoted $\EF'_{i}$ (see Appendix \hyperref[sec:AppArbitAngle]{C}).
The Rabi frequency becomes
\begin{equation}
\Omega(\Theta, m_J')= \sum_{i, m'} \frac{\EF'_{i} \, \beta'_{m'} }{\hbar}
\langle e_k (r), m_J' \vert (-e) r'_{e,i} \vert g, m' \rangle \,.
\end{equation}
 Clearly, $\Omega$ depends on $m_J'$, the magnetic quantum number of the excited state in the molecular frame, the PEC with index $k$, and the internuclear separation $r$, all of which are essentially determined in the experiment by the excitation laser frequency.
 We also see that, in contrast to our previous assumption in section~\ref{sec:simplescheme}, $\Omega$ does in general depend on the angle $\Theta$.
 Furthermore, when calculating $\Omega$ we have to take into account that
 the Rydberg states (PECs) come in degenerate pairs of $\vert \pm m_J' \rangle$.
  Therefore, the effective optical coupling $\Omega$ into the excited manifold spanned by $\vert \pm m_J' \rangle$   is given by the quadrature sum of the corresponding Rabi frequencies, i.e.
  \begin{equation}
  \Omega(\Theta)  = \sqrt{\Omega^2(\Theta,m_J') + \Omega^2(\Theta,-m_J')}\,.
  \label{eq:quadsum}
\end{equation}
For further discussion see Appendix~\hyperref[sec:AppArbitAngle]{C}.
Figure~\ref{figAngleDependence} shows a few examples for $\Omega (\Theta)$ for various light polarizations and $m_J'$. The coupling-laser intensity is $I_L = 400\:$mW/$(\upmu \text{m}^2)$.
We use the dipole matrix elements $d_i(e_k)$ at the internuclear separations $r$
that correspond to the local maxima of the PECs for the avoided crossings $C_I$ - $C_{III}$  in Fig.~\ref{fig:PotCurvesDipoleMoments}.
 These crossings  can potentially be used for ion-atom shielding, which will be explained in detail in the next section.
As a simple rule, shielding will work better the  larger $\Omega$ is and  the more isotropic $\Omega$ is.
\begin{figure}
	\centering
	\includegraphics[width=\columnwidth] {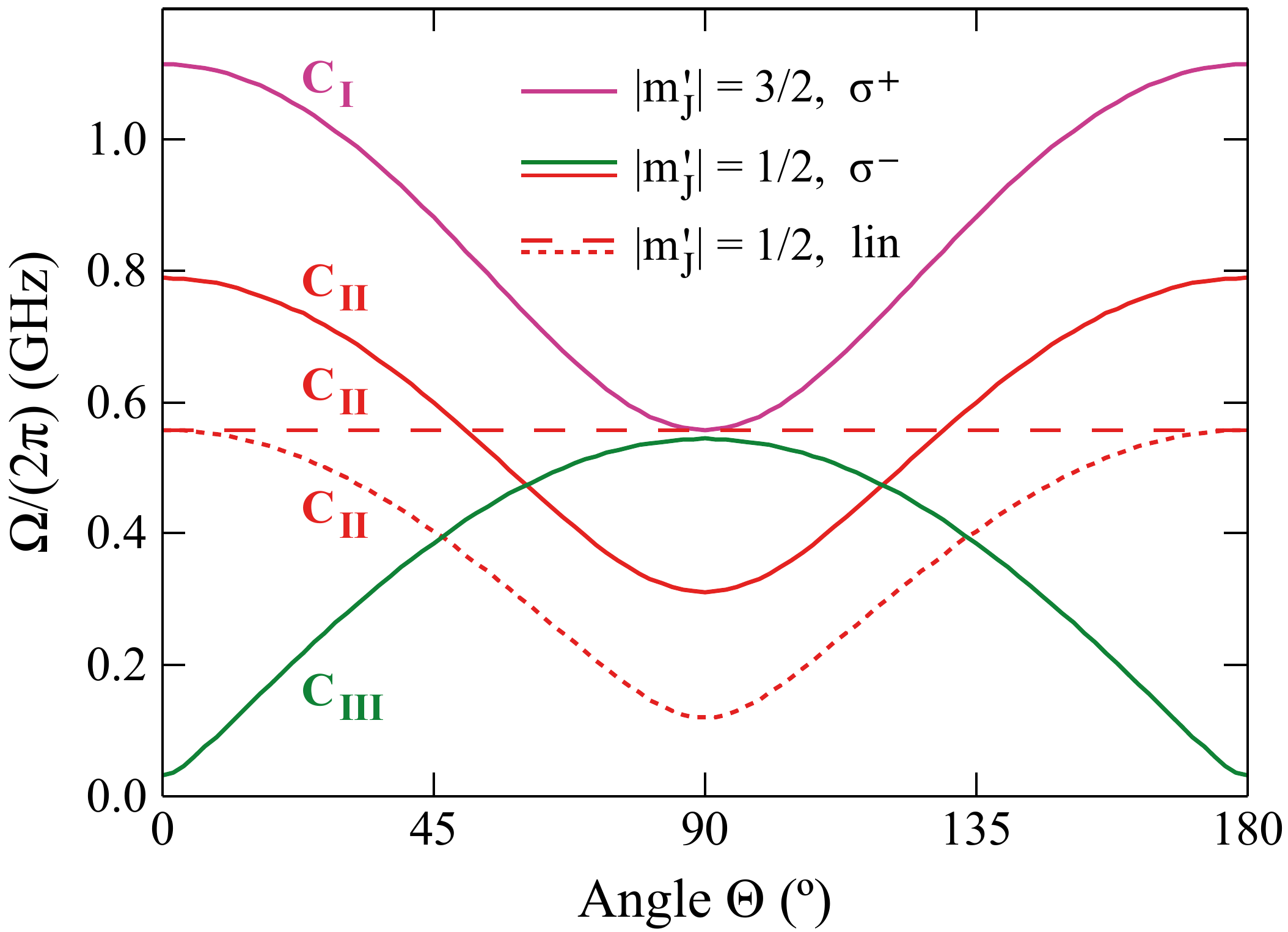}
	\caption{Calculated angular dependence of the Rabi frequency $\Omega$ at the avoided crossings marked $C_{I}, C_{II}$ or $C_{III}$ in Fig.~\ref{fig:PotCurvesDipoleMoments}. The atoms are prepared in the ground state $|5S_{1/2}, m= 1/2 \rangle$ in the lab frame.
The laser intensity is $I_L = 400\:$mW/$(\upmu \text{m}^2)$. The indicated polarizations are relative to the lab-frame's $z$-axis. $|m_J'|$ specifies the addressed Rydberg manifold  $| e_k, \pm m_J'  \rangle$ in the molecular frame.
The internuclear separations $r$ in the calculation correspond to the PEC maxima at the $C_{I}, C_{II}$ and $C_{III}$ avoided crossings, and are 76.8~nm, 73.7~nm and 83.4~nm, respectively.
	The dashed and dotted red lines are for a  linearly polarized laser.
  The red dashed line shows the case when the polarization points in $y$-direction, while for
  the red dotted line it points in $x$-direction. 
	}
	\label{figAngleDependence}
\end{figure}
Generally, we note the symmetry of the curves in Fig.~\ref{figAngleDependence}  about $\Theta = \pi/2$, which is due to the fact that $\Omega(\Theta)$ has the form of Eq.~(\ref{eq:quadsum}).
The magenta solid line is for coupling to the $C_I$-crossing with a circularly polarized laser.
For $\Theta = 0$, the laser drives a $\sigma^+$-transition from  $|g, m = 1/2 \rangle$ to $|e, m_J = 3/2 \rangle$ with a  Rabi frequency $\Omega = 2\pi \times 1.11$~GHz, which, as we will see in the next section, is sufficient for successful shielding. At $\Theta = 90^\circ$, $\Omega$ drops to half of its value at $\Theta = 0$.
This can be mainly explained by the fact that at $\Theta = 90^\circ$ a sizable fraction of the light is $\pi$-polarized
in the molecular frame, and $\pi$-polarized light cannot  drive a transition from the $J = 1/2$ ground state  to the  $m_J = 3/2$ Rydberg state.

 The red continuous line is for the crossing $C_{II}$, where the excited state is a $| m_J'| = 1/2$ manifold.
 We use a circularly polarized laser driving a  $\sigma^-$-transition at $\Theta = 0$.
 Overall, the coupling $\Omega$ is weaker than for $C_{I}$, but it exhibits similar angular dependence. The relative loss at $\Theta = 90^\circ$ is slightly stronger than for $C_{I}$. The loss is due to an interference effect where  $\sigma$ and  $\pi$  transition paths from the ground state in the rotated frame, $|g \rangle =   (|g, m' = 1/2\rangle - |g, m' = -1/2\rangle)/ \sqrt{2}$, to either $|e_k, m_J' = \pm 1/2 \rangle$ destructively interfere.
In order to make the coupling more isotropic we now try using  linear polarization  (dashed and dotted red lines) instead of circular polarization. 
The polarization direction of the laser light now breaks the rotational symmetry about the $z$-axis
which exists for circularly polarized light.
Therefore,  we now analyze the  dependence on the angle $\Theta$ for two cases: 1)  polarization in $y$-direction (red dashed line) and 2) polarization in $x$-direction (red dotted line). The dashed line is flat and therefore $\Omega$ is independent on $\Theta$ with respect to rotation about the $y$-axis. The dotted line, however, exhibits an increased anisotropy.
 At angle $\Theta = 0$ the Rabi frequency $\Omega$ for the dashed and dotted lines is by a factor of  $1/\sqrt{2}$ smaller than for circular polarization (red solid line),  because only one of the two circular components of the linear polarization contributes to the coupling towards the $|m_J| = 1/2$ Rydberg manifold.

The green line in Fig.$\:$\ref{figAngleDependence} is for  $\sigma^-$ laser light at $\Theta = 0$, coupling to avoided crossing $C_{III}$. From Fig.~\ref{fig:PotCurvesDipoleMoments} we gather that for this case only $\pi$-transitions have sizable transition moments $d_i(e_k)$. Therefore, at   $\Theta = 0$, $\Omega$ almost vanishes.
At $\Theta = 90^\circ$  where the light polarization  has a strong $\pi$-component in the rotated frame the coupling is maximal.

Hence, among the examples studied the crossing $C_I$ with a $\sigma^+$ drive is the best choice,
since it has the strongest overall coupling and a comparatively small angular anisotropy.

For completeness, it needs to be checked whether coupling to other PECs is negligible because this can lead to complications and losses.
For example, the $C_I$ crossing is energetically  close to the avoided crossing $C_{IV}$, as can be seen in Fig.~\ref{fig:PotCurvesDipoleMoments}, which also exhibits sizable transition matrix elements. However, closer inspection shows that the two potential curves  are still split by about $h \times 6$~GHz, which should be enough to treat them separately.

We note, however, that coupling to other PECs can also be an interesting feature.
For this, we consider crossing $C_{IV}$ in Fig.~\ref{fig:PotCurvesDipoleMoments}. In order
to reach $C_{IV}$ an atom would have to cross the $17P_{1/2}$ level (black dashed line) at a distance of about 90$\:$nm.
At this point some incoming flux can be coupled to the $17P_{1/2}$ level, which will
continue following the dashed PEC, adiabatically pass the crossing $C_{III}$, and will be reflected on the inward up-slope of the dashed PEC at $r \sim 80$~nm. On exit, i.e. at the crossing at 90~nm, the wave on the dashed PEC will interfere with the (partially) reflected wave from the crossing $C_{IV}$. This opens up the possibility to interferometrically control the total reflection probability or to tune the phase of the net reflected wave. We can estimate that the coupling to the $17P_{1/2}$ PEC can be sizable by using Eq.~(\ref{eq:OmegaBar}), $\Omega \approx 2\pi \times 1$ GHz  and  that the slope of the PEC at $r \sim90$~nm is $dV_e/ dr \approx 8$ GHz/nm$\times h$. Further exploration of this topic is beyond the scope of the present paper.

To summarize,  we find that for certain avoided crossings and coupling-laser polarizations the optical coupling $\Omega(\Theta)$ varies by less than a factor of 2 to 3, still leading to robust shielding 
for a sufficiently large optical coupling $\Omega$.
As briefly mentioned before in section \ref{sec:CaseStudyA}, the angular dependence in  $\Omega$ will lead to an anisotropic effective scattering potential between atom and ion. This will cause mixing of different partial waves. However, such a mixing is not of interest here and beyond the scope of this work.
Therefore, in the following we will carry out calculations where we ignore the angular dependence of $\Omega$.

\section{Numerical solution for shielding}

We now investigate the collision dynamics by numerically solving the Schr\"{o}dinger equation for the scattering problem. For this, we consider only two coupled channels: (1) The atomic  ground state with potential energy $V_g(r)$, and (2) the Rydberg state with potential energy $V_e(r)$. The two channels are coupled via a laser with the Rabi frequency $\Omega(r)=d_i(r) \times \sqrt{2I_L/(c\varepsilon_0)}/\hbar$, which depends on the relative distance of the atom and the ion. Here, $c$ is the speed of light. 
As discussed at the end of  section \ref{sec:TransDipole} we assume here the coupling strength to be spherically symmetric. Within the rotating wave approximation
the coupled potentials for the two-level system can be written as
\begin{equation}
V_I=
\left( \begin{array}{cc}
V_{g}(r)  &  \hbar\Omega(r)/2 \\
\hbar\Omega(r)/2	  &   V_{e}(r)
\end{array}\right)\,,
\label{eq:VIfull}
\end{equation}
generalizing Eq.$\:$(\ref{eq:VIlinear}). With the new $V_I$ we  numerically solve the Schr\"{o}dinger equation Eq.$\:$(\ref{eq:SE}) using the same method as before, see sections \hyperref[sec:NumSolution]{A} and \hyperref[sec:AdiabaticBasis]{B} of the Appendix. Here, the two components of the wave function are denoted by $\varphi_U$ and $\varphi_L$, where the subscript indicates the energetically lower branch ($L$) and the energetically upper branch ($U$), respectively. The task is to calculate scattering solutions for flux entering from large internuclear distance in channel $U$. This flux can then be reflected, transmitted, or it can non-adiabatically leak into channel $L$.

\begin{figure}
	\centering
	\includegraphics[width=\columnwidth] {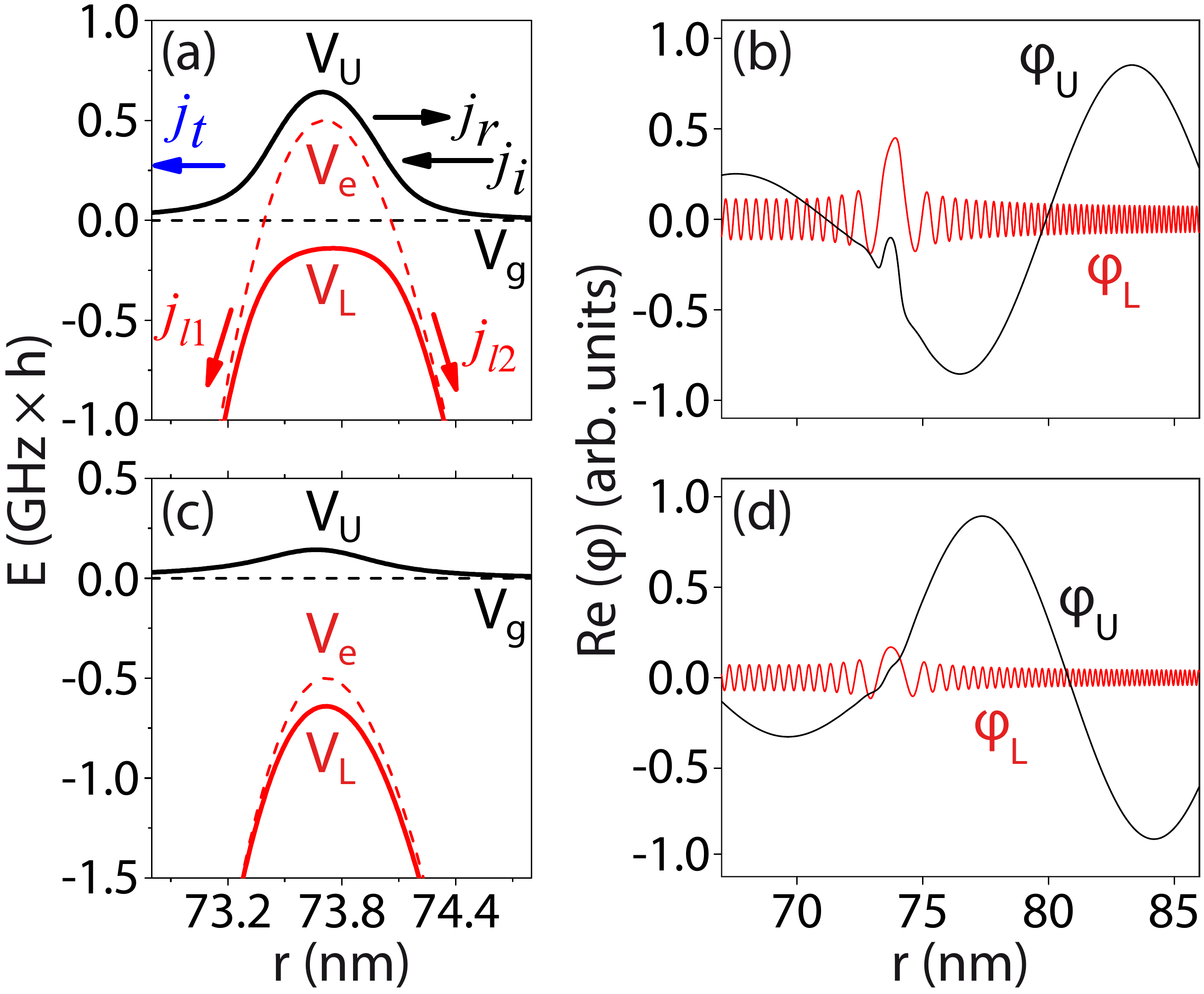}
	\caption{(a) Calculated potential energies with coupling ($V_U$ and $V_L$, solid lines) and without coupling ($V_e$ and $V_g$, dashed lines) for $\Delta=2\pi \times -500\:\text{MHz}$, $\Omega=2\pi \times 730\:\text{MHz}$  at the peak of $V_U$, and $E_{coll}=1\:\text{mK}\times k_\text{B}$.
	$j_i, j_r, j_t$ are the incoming, reflected and transmitted currents, respectively. $j_{l1}$ and
	$j_{l2}$ are the currents that non-adiabatically leak into channel $V_L$. (b) Real parts of the calculated scattering wave functions $\varphi_U$ (black solid line) and $\varphi_L$ (red solid line). (c) and (d) are calculations using the same parameters as in (a) and (b), except that $\Delta=2\pi \times 500\:\text{MHz}$.
}
	\label{fig3}
\end{figure}

Figure \ref{fig3}(a) shows an example for the resulting potential curves $V_U$, $V_L$ (solid lines) as well as the uncoupled energies $V_g$, $V_e$ (dashed lines), corresponding to the avoided crossing $C_{II}$ of Fig.$\:$\ref{fig:PotCurvesDipoleMoments}(b). The laser has a detuning of $\Delta= -2\pi \times500\:\text{MHz}$ from the
tip of $V_e$, i.e. the tip of $C_{II}$.
 Furthermore, we choose a laser intensity of 343$\:$mW/($\upmu$m)$^2$ which corresponds to a maximal Rabi frequency of $\Omega =2\pi  \times730\:\text{MHz}$ at $r=73.7\:\text{nm}$. We choose a collision energy of $E_{coll}=1\:\text{mK}\times k_\text{B}$ which is defined at $r \rightarrow \infty$. Figure \ref{fig3}(c) shows the potential energy curves for identical conditions except that $\Delta=2\pi\times 500\:\text{MHz}$.
 While in Fig.~\ref{fig3}(a) there are two avoided crossings between $V_g$ and $V_e$,  there is no crossing
  in Fig.~\ref{fig3}(c).

The numerical solutions (real parts) for the scattering wave functions  for the potential curves in Figs.$\:$\ref{fig3}(a) and (c) are shown in plots (b) and (d), respectively. The distortion of the wave functions  $\varphi_U$ at around $r = 74\:$nm indicates that the non-adiabatic coupling takes place only in the vicinity of the avoided crossing, as expected. Furthermore, it can be clearly seen that the amplitude of $\varphi_U$ is much smaller to the left of the barrier than to the right, indicating efficient shielding.

 \begin{table}
	\begin{tabular}{|l|c | c|}
		\hline
		current ratio & case (a) & case (c) \\
		\hline
		$j_r / j_i$ &   52\%    &  71\%    \\
		$j_t / j_i$ &   3\%    &  7\%    \\
		$(j_{l1} + j_{l2}) / j_i$ &   45\%    &  22\%    \\
		\hline
	\end{tabular}
	\caption{Ratios of probability currents for the scattering solutions in Fig.\:\ref{fig3}. Case (a)  refers to
 Fig.\:\ref{fig3}(a) and (b),  while case (c) refers to  Fig.\:\ref{fig3}(c) and (d).}\label{table1}
\end{table}

We quantify the reflection by comparing incoming, reflected and leaking probability currents. For this, we choose locations $r$ far away from the barrier and express the scattering wave functions for each scattering channel $q \in \{ U, L \} $ in terms of $\varphi_q=a_q(r) e^{ik_qr} + b_q(r) e^{-ik_qr}$, as described in section \ref{sec:simplescheme}. The outward and inward currents are $j_{out,q}(r) = |a_q(r)|^2 k_q\hbar/\mu$ and $j_{in,q} = |b_q(r)|^2 k_q\hbar/\mu$, respectively, with $k_q=\sqrt{2\mu (E_{coll}-V_q(r))}/ \hbar$. As shown in Fig.$\:$\ref{fig3}(a), we label the incident current as $j_i = j_{in, U}(r > 76\:\text{nm})$, the reflected current as $j_r = j_{out, U}(r > 76\:\text{nm})$, the transmitted current as $j_t = j_{in, U}(r < 72\:\text{nm})$, and the currents corresponding to non-adiabatic leakage into the $L$-channel as $j_{l1} = j_{in, L}(r < 72\:\text{nm})$ and $j_{l2} = j_{out, L}(r > 76\:\text{nm})$.

 \begin{figure*}
 	\centering
 	\includegraphics[width=1.5\columnwidth] {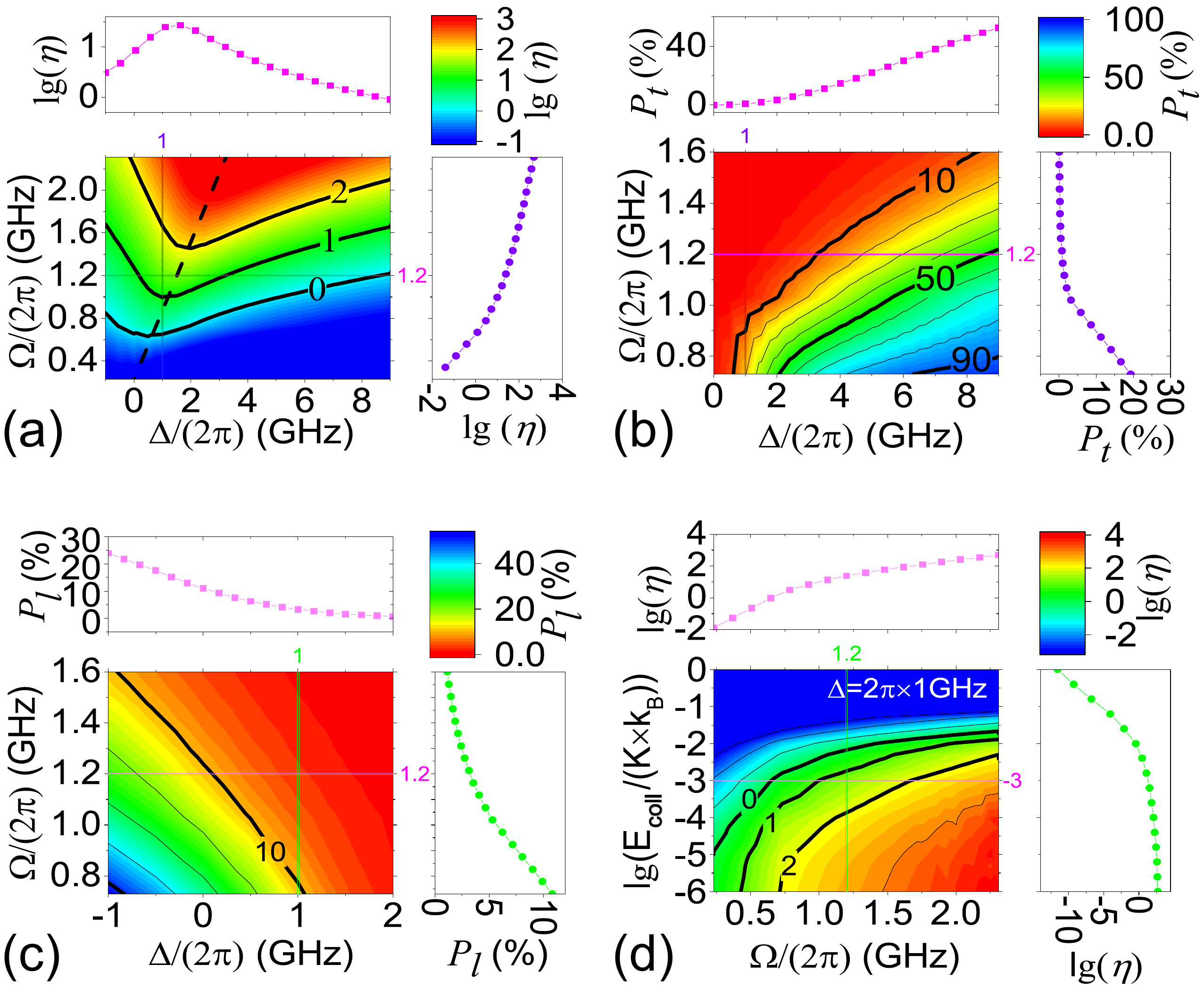}
 	\caption{(a) Shielding efficiency $\eta$ as a function of laser detuning $\Delta$ and coupling strength $\Omega$ for a collision energy of $1\:\text{mK}\times k_\text{B}$. The dashed black line indicates the best choice of $\Delta$ to reach optimal shielding for a given $\Omega$. (b) and (c) show the transmitted fraction $P_t$ and the non-adiabatic leaking fraction $P_l$   versus $\Delta$ and $\Omega$. (d) shows $\eta$ versus  $\Omega$ and collision energy $E_{coll}$. $E_{coll}$ is in units of $\text{K}\times k_\text{B}$. $\Delta$ = $2\pi \times 1\:$GHz. $\lg \equiv \log_{10}$ is the decadic logarithm. The plots above and on the right hand side of each contour plot are cuts as indicated by the corresponding lines in the contour plots.
 	}
 	\label{Fig.4}
 \end{figure*}

In Table \ref{table1} we list the reflection, transmission and adiabatic-loss percentages for the two examples in Figs.$\:$\ref{fig3}(a) and (c). For the sake of the discussion, the parameters for the two examples have been chosen such that there are still  sizable tunneling and leakage currents. As becomes clear from Table \ref{table1}, the case $\Delta > 0$ yields better shielding. We will show further below that the shielding efficiency can be nearly 100\% when globally increasing the coupling strength $\Omega$ by a factor of 2, e.g. by increasing the laser power by a factor of 4.

Besides the reflection probability $P_r = j_r/j_i$ it is convenient to define a second measure, $\eta$, for the shielding efficiency
\begin{equation}
\eta = j_r / (j_t+j_{l1}+j_{l2}) = j_r / (j_i - j_r) \,,
\end{equation}
which gives the ratio of (good) reflected flux to (bad)  flux  lost in unwanted channels.
Figure \ref{Fig.4}(a) shows $\log_{10}(\eta)$ as a function of $\Delta$ and $\Omega$ for a collision energy of $1\:\text{mK}\times k_\text{B}$. The solid black contours labeled by `0', `1' and `2' correspond to a reflection probability of $P_r$ = 50$\%$, 91$\%$ and 99$\%$, respectively. Quite generally, the shielding efficiency $\eta$ increases with coupling strength $\Omega$. Furthermore, for a given  $\Omega$, shielding is best for $\Delta \approx \Omega$ (see dashed black line).
If $\Delta$ becomes too large, the repulsive barrier becomes so small, that either strong tunneling through the barrier occurs or flux even passes over the barrier. Figure \ref{Fig.4}(b) illustrates this in a  plot of the transmission probability $P_t = j_t/ j_i$.  $P_t$ increases with $\Delta$ and decreases with $\Omega$. We have numerically checked that the transmission probability function $P_t(\Omega, \Delta)$ shown in Fig.$\:$\ref{Fig.4}(b) can be approximately reproduced in the shown range with the well-known expression for 1D-tunneling
\begin{equation}
P_t(\Omega, \Delta)= \exp\left( -2\frac{\sqrt{2\mu}}{\hbar}\int_{Tp1}^{Tp2} \sqrt{V_U(r) - E_{coll}} \, dr \right)\,,
\end{equation}
where $T_{p1}$ and $T_{p2}$ are the classical turning points of the potential barrier $V_U(r)$. This means that within the shown parameter range, the transmission is dominated by tunneling, and non-adiabatic transitions onto $V_L$ only play a minor role. For $\Delta > 0$, we find that the tunneling barrier $V_U$ is approximately described by a Lorentzian profile,
\begin{equation}
V_U(r)\approx V_0 \frac{( \Gamma/2)^2}{( \Gamma/2)^2+ (r-r_s)^2}\,,
\end{equation}
with height $V_0 = 0.5 (-\Delta + \sqrt{ \Omega^2 + \Delta^2)} \hbar$ and width $ \Gamma = \sqrt{2 \Omega^2 \, \hbar / (V_0 \, k_{H}) } $. Here, $k_H$ is the negative curvature of the potential curve $V_e$ at its local maximum. For tunneling through such a Lorentzian barrier analytical results for the transmission probability $P_t$ can be derived, as described in section \hyperref[sec:AppendixTunnelling]{D} of the Appendix.

Next, we discuss non-adiabatic transitions which leak flux to the lower potential energy curve $V_L$. Figure \ref{Fig.4}(c) shows the probability for this leakage $P_l = (j_{l1} + j_{l2})/j_i$  as a function of $\Omega$ and $\Delta$. In the range shown, $P_l$ decreases monotonically with increasing $\Omega$ and $\Delta$. While it is plausible that a larger $\Omega$ generally improves the adiabaticity, we note that for vanishing $\Omega$ the leakage also will vanish because the PECs $V_e$ and $V_g$ are not coupled anymore. However, a vanishing $\Omega$ is not of interest for our discussion here. The dependence of $P_l$ on $\Delta$ in Fig.$\:$\ref{Fig.4}(c) can be understood as follows. For $\Delta < 0 $ a decrease of $|\Delta|$ decreases the slope $dV_e/dr$ at the crossing, which increases the adiabaticity according to our discussion in section \ref{sec:simplescheme}. For $\Delta > 0 $ there is  an inherent  momentum mismatch for coupling flux from the  upper  to the lower channel, which suppresses non-adiabaticity. The mismatch and therefore the adiabaticity increase with $\Delta$.

From the discussion in section \ref{sec:simplescheme} where the avoided crossing with a linear potential energy curve is studied one might expect that for $\Delta < 0$ the leakage $P_l$ is a Landau-Zener-like function of the scaled quantity  $\bar{\Omega}$. However, this is only valid for $\Omega \ll | \Delta | $ and small enough $E_{coll}$. In the parameter range discussed here, we find that $P_l$ roughly scales as
\begin{equation}
P_{l} \approx \exp( c_1 + c_2 \, \hbar \Omega / E_s  + c_3 \, \hbar \Delta / E_s)\,,
\label{eq:Pl}
\end{equation}
where  $E_s = 0.5 \hbar \sqrt{k_H / \mu}$ is an energy scale as determined by the negative curvature $ k_H = -d^2V_e/dr^2$ of the barrier at its peak, see section \hyperref[sec:AppendixTunnelling]{D} of the Appendix for details. The coefficients $c_i$ vary slowly with $\Omega$ and $\Delta$. For a small enough collision energy $E_{coll}$, $P_{l}$ scales like a power law, i.e. $ P_{l} \propto (E_{coll}/E_s) ^{\alpha}$. Here, $\alpha$ is a slowly varying function of $\hbar \Omega / E_s$ and  $\hbar \Delta / E_s$.  This means that the coefficients $c_i$ in Eq.~(\ref{eq:Pl}) can  be expanded as
\begin{equation}
c_i = c_{i,1} + c_{i,2} \ln(E_{coll}/ E_s)\,.
\end{equation}

Figure \ref{Fig.4}(d) shows the shielding efficiency $\eta$  versus the initial collision energy $E_{coll}$ and $\Omega$. Here, $\Delta$ is set to $\Delta = 2\pi \times 1\:$GHz. As expected, shielding improves as the collision energy is lowered, because both tunneling and non-adiabaticity are increasingly suppressed.

Similar as for the non-adiabatic leakage, $\eta$ exhibits approximately power law scaling, $\eta \propto (E_{coll}/ E_s)^\alpha$, as long as the collision energies $E_{coll}$ are small enough. As before, the exponent $\alpha$ depends on $\hbar \Omega / E_s$ and $\hbar \Delta / E_s$.

\section{Conclusion}

In conclusion, we propose a method for  shielding a cold neutral  atom and an  ion from a collision at close range. When the particles reach an interparticle distance on the order of 100~nm  the neutral atom is resonantly excited to a low-field-seeking Rydberg level which is repelled by the ion. Upon leaving  the neutral atom is de-excited back in an adiabatic way, so that no spontaneous scattering of photons occurs. We find that this shielding scheme is particularly interesting  when employing an avoided crossing of two Rydberg levels. We discuss how shielding  depends on the Rabi frequency of the laser, on the laser detuning from the avoided crossing of a Rydberg level,  on the collision energy, and on the collision angle. At collision energies of about $1\:\text{mK}\times k_\text{B}$ typically Rabi frequencies on the order of $\Omega = 2\pi \times1\:$GHz are needed for efficient shielding. The shielding efficiency can be varied from zero to nearly 100\% by adjusting the laser intensity and frequency. In future work one may investigate the coupling between different partial waves caused by the anisotropy of the shielding potential, as well as matter-wave interference between multiple avoided crossings as a method for collision control.

\section*{ACKNOWLEDGMENTS}

We gratefully acknowledge funding support by DFG Priority Programme 1929. We would like to thank Guido Pupillo and Nora Sandor for helpful discussions.

\section*{APPENDIX}

\subsection{Numerical solution of the Schr\"{o}dinger equation}
\label{sec:NumSolution}
Here, we describe how we  determine the scattering solution of the Schr\"{o}dinger equation Eq.$\:$(\ref{eq:SE}) for the interaction potential of Eq.$\:$(\ref{eq:VIfull}). We are looking for a particular solution for which ground state atom and ion approach each other with collision energy $E_{coll}$ (defined at $r = \infty$). After switching to the adiabatic basis (see section \hyperref[sec:AdiabaticBasis]{B} of the Appendix) the Schr\"{o}dinger equation is numerically integrated starting from a suitable position $r_0$ on the left of the avoided crossing towards increasing $r$. The position $r_0$ is chosen to be sufficiently far away from the avoided crossing such that the coupling of the ground state channel and the Rydberg channel is negligible. According to the  boundary condition of having the incoming wave in the ground state and approaching from $r = \infty$, the wave function components $\varphi_U$ and  $\varphi_L$ at $r_0$ must be outgoing with respect to the avoided crossing, i.e.  $\propto \exp(-i k_q r)$ with  the local wavenumber $k_q = \sqrt{2\mu(E_{coll}-V_{q}(r_0))}/ \hbar$ for  $q\in\{ U, L\}$. This determines the derivatives of the wave function components at this point to be $d\varphi_q/ dr = -i k_q \varphi_q$. We separately carry out two integrations with two linearly independent starting vectors $\left( \varphi_U(r_0),  \varphi_L(r_0) \right)^T$. Afterwards, the two solutions are linearly combined to provide the desired final solution which fits the boundary condition.

Finding the scattering solution for the interaction potential of Eq.$\:$(\ref{eq:VIlinear}) is analogous, apart from setting  $d\varphi_U/ dr(r_0) = k_U \varphi_U(r_0)$ with
$k_U = \sqrt{2\mu(E_{U}(r_0) - E_{coll}) }/ \hbar$. After the numerical solution of the Schr\"{o}dinger equation the wave functions can be expressed again in the non-adiabatic basis $|g \rangle $ and $|e \rangle $ as described in section \hyperref[sec:AdiabaticBasis]{B} of the Appendix.

\subsection{Adiabatic basis for solving the Schr\"{o}dinger equation}
\label{sec:AdiabaticBasis}
In order to numerically integrate the Schr\"{o}dinger equation Eq.$\:$(\ref{eq:SE}) it can be advantageous from a numerical point of view to locally express the two-component wave function $\Psi  = (\varphi_{g}, \varphi_{e})^T$ in a basis for which the interaction Hamiltonian $V_I$ is diagonal. This is done by the following transformation $\hat S^{-1} \Psi  = \Phi \equiv (\varphi_U, \varphi_L)^{T}$ (see \cite{Kazantseu1990}), where $\hat S$ diagonalizes $V_I$ by
\begin{equation}
\hat{S}^{-1} V_I \hat{S} =
\left( \begin{array}{cc}
V_{U}(r)  &  0 \\
0	  &   V_{L}(r)
\end{array}\right)\,.
\end{equation}
We note that $\hat{S}$ is unitary, i.e. $\hat{S}^{-1} = \hat{S}^\dagger$. This basis change transforms the Schr\"{o}dinger equation into
\begin{equation}
\left( -\frac{\hbar^2}{2\mu} \frac{d^2}{dr^2} + \hat R  + \left( \begin{array}{cc}
V_{U}(r)  &  0 \\
0	  &   V_{L}(r)
\end{array}\right) \right)  \Phi  = E_{coll}  \Phi \,,
\end{equation}
where $\hat{R}=\hat{R}_{1}+\hat{R}_{2}$ is the non-adiabaticity operator with
\begin{eqnarray}
\hat{R}_{1} & = &-i\frac{\hbar^2}{\mu}\hat{S}^{-1} \left( \frac{d\hat{S}}{dr}\right)  \frac{d}{dr}\,, \\
\hat{R}_{2} &=& -\frac{\hbar^2}{2\mu}\hat{S}^{-1} \left( \frac{d^2\hat{S}}{dr^2} \right)\,.
\end{eqnarray}
Expressing $\hat R_{1}$ and $\hat R_{2}$ in terms of  matrices,
\begin{eqnarray}
\hat{R}_{1} & = & \left(  \begin{array}{cc}
R_{1d}   &  R_{1nd} \\
-R_{1nd}	  &  R_{1d}\
\end{array}
\right) \frac{d}{dr}\,,\\
\hat{R}_{2} &=& \left(
\begin{array}{cc}
R_{2d}   &  R_{2nd} \\
-R_{2nd}	  &   R_{2d}\
\end{array}
\right)\,,
\end{eqnarray}
we obtain the coupled Schr\"{o}dinger equation in the following form,
\begin{eqnarray}
\frac{d^2 \varphi_{U}}{dr^2} &=& -\frac{2\mu}{\hbar^2}[(E_{coll} -V_U-R_{2d})\varphi_U-R_{1d}\frac{d \varphi_{U}}{dr} \nonumber \\ & &  -R_{1nd}\frac{d\varphi_{L}}{dr}- R_{2nd}\varphi_L]\,,\\
\frac{d^2 \varphi_{L}}{dr^2} & = & -\frac{2\mu}{\hbar^2} [ (E_{coll} -V_L-R_{2d})\varphi_L+R_{1nd}\frac{d \varphi_{U}}{dr} \nonumber \\ & & + R_{2nd}\varphi_U -R_{1d}\frac{d\varphi_{L}}{dr} ]\,.
\end{eqnarray}
The non-diagonal elements of $\hat{R}$  mix the channels $U$ and $L$. They are only appreciable close to the avoided crossing.

\subsection{Rabi frequency $\Omega$ for arbitrary collision angles }
\label{sec:AppArbitAngle}
Here, we calculate the Rabi frequency  $\Omega$ along with the effective transition dipole matrix element $ d_i(e_k) $ for an optical transition from the atomic ground state to a Rydberg state for the case when the quantization axis $z$ is not collinear with the internuclear axis $z'$ of atom and ion.
Let the ground state atom be in the state $|g\rangle = \sum_{m} \beta_m |g, m \rangle$. Here, the angular momentum $J = 1/2$ is not indicated. The electrical field of the laser is $ \vec{\EF} = (\EF_{x}, \EF_{y},\EF_{z})$. The light field can be  decomposed into the spherical components:
 \begin{eqnarray}
 \EF_1 &=& - { 1 \over \sqrt{2} }  (\EF_x - i \EF_y) = \EF \, a_1 \,,\\
 \EF_0 &=&  \EF_z  = \EF \, a_0 \,, \\
 \EF_{-1} &=& { 1 \over \sqrt{2}}  (\EF_x + i \EF_y) = \EF \, a_{-1}\,,
 \end{eqnarray}
 where $\EF = \sqrt{\EF_x^2 + \EF_y^2  + \EF_z^2 }$ and the  \{$a_i$\} are relative amplitudes of the light polarization components ($\sum_i \vert a_i \vert^2 =1$).

 The rotation from the lab frame $\{x,y,z\}$  into the molecular frame $\{x',y',z'\}$, for which the internuclear axis $z'$ is the quantization axis, is effected by a rotation vector $\vec{\Theta}$ with magnitude $\Theta$ (see also Fig.~\ref{Fig.1}).
 For simplicity the frames are chosen such that the $y,y'$-axes point along $\vec{\Theta}$.
The atomic ground state in the molecular frame is
 \begin{eqnarray}
 |g \rangle =  \sum_{m,m'} d_{m',m}^{(1/2)}(-\Theta) \beta_{m} |g, m' \rangle\,.
 \end{eqnarray}
 Here, $d_{m',m}^{(1/2)}(-\Theta)$ is given via $d_{\tilde{m},m}^{(1/2)}(\Theta) = \langle g, \tilde{m}| \exp(-i \Theta \, J_y) |g,m \rangle$
representing Wigner's (real-valued) small d-matrix  for $J = 1/2$. $J_y$ is the $y$-component of the angular momentum operator.
We would like to point out, that throughout the paper  a prime ($'$) inside a bra or ket, e.g. $|g, m' \rangle$, has a double meaning: a) it creates a new variable name (here, $m'$) and b) it indicates that the quantum numbers are determined in the molecular frame.  Bras or kets without prime are in the lab frame.
   The  polarization amplitudes $\{ a_i \}$ of the light in the lab frame transform into
   $\{ a'_i \}$ for the primed-frame, 
\begin{equation}
a'_i = \sum_{k} d^{(1)}_{i,k}(-\Theta) a_k \, ,
\label{eq:ai}
\end{equation}
where the rotation matrix elements are for $J=1$ and the rotation also is about the $y$-axis.

  The atom-light interaction Hamiltonian is given by
 \begin{eqnarray}
 H_{AL} &=& -e \vec{r}_e \cdot \vec{\EF}  \nonumber \\
 &=& -e   \sum_i r_{e,i} \, \EF_i     \nonumber \\
 &=& -e \EF  \sum_i r_{e,i} \, a_i\,,
 \end{eqnarray}
 where
 \begin{eqnarray}
r_{e,1} &=& - { 1 \over \sqrt{2} }  (x_e + i y_e)\,, \\
r_{e,0} &=&  z_e \,, \\
r_{e,-1} &=& { 1 \over \sqrt{2}}  (x_e - i y_e)\,,
\end{eqnarray}
are the spherical components of $\vec{r}_e$. The components $r_{e,1}, \, r_{e,0}, \, r_{e,-1}$ can be viewed as operators which induce $\sigma^+, \pi, \sigma^-$ transitions, respectively, so that the $m$-quantum number of the atom changes by $1, 0, -1$, respectively.

In the rotated coordinate system $\{x',y',z'\}$ the Hamiltonian for atom-light interaction reads
\begin{eqnarray}
 H_{AL}  &=& -e \EF  \sum_i r'_{e,i} \,  a'_i\,,
 \end{eqnarray}
 with polarization amplitudes $\{ a'_i \}$ as defined in Eq.~(\ref{eq:ai}).
 The ket for a molecular Rydberg state with angular momentum $m_J'\hbar$
about the internuclear axis is $\vert e_k, m_J' \rangle$, where the label $k$ specifies the  potential energy curve. We note that also the internuclear separation $r$ is implicitly fixed.
 The transition matrix element from the ground state $| g \rangle$ (which is a 5S$_{1/2}$ state) to the Rydberg state $| e_k, m_J' \rangle$
 is given by
 \begin{eqnarray}
& & \langle  e_k, m_J'| H_{AL} | g \rangle \nonumber  \\
  &=&  \EF  \langle  e_k, m_J'|   \sum_{i,n}   (-e) r'_{e,i}  d_{i,n}^{(1)}(-\Theta)  a_n | g \rangle  \nonumber \\
  &=& \EF   \sum_{i,m', m, n} \langle e_k, m_J' |  (-e) r'_{e,i}  |g,  m' \rangle \times \\
  & &  \quad  d_{i,n}^{(1)}(-\Theta) \,  a_n \,\,  d_{m',m}^{(1/2)}(-\Theta)  \, \beta_m \,.
 \end{eqnarray}

 The term $ \langle e_k, m_J' |  (-e) r_{e,i}'  |g,  m' \rangle  =  d_{i}(e_k)$ is
 the electric dipole transition matrix element, as defined in Eq.~(\ref{eq:diek}) and calculated  earlier
 in section \ref{sec:TransDipole}.
 We note that due to rotational invariance,
 $ \langle e_k, m_J' | r_{e,i}'  |g,  m' \rangle  = \langle e_k, m_J | r_{e,i}  |g,  m \rangle $, when
 $m_J' = m_J$ and  $m' = m$.
The Rabi frequency $\Omega (m_J')$ is given by
 \begin{eqnarray}
 \Omega (m_J')  \equiv \langle  e_k, m_J'| H_{AL} | g, m \rangle   / \hbar \,.
 \end{eqnarray}
Due to the degeneracy of the Rydberg states $ \vert e_k, m_J' \rangle $ and $ \vert e_k, - m_J' \rangle $
the effective
optical coupling strength is
\begin{equation}
\Omega = \sqrt{\Omega^2 ( m_J') + \Omega^2 (- m_J')}\,,
\end{equation}
coupling to the superposition state
\begin{eqnarray}
  {\Omega_{m_J'} | e_k, m_J' \rangle +  \Omega_{-m_J'}| e_k, -m_J' \rangle  \over \sqrt{\Omega_{m_J'}^2 +\Omega_{-m_J'}^2} }\,.
 \end{eqnarray}

\subsection{Approximate analytic expression for the tunneling amplitude}
\label{sec:AppendixTunnelling}
Here, we derive an approximate, analytical expression for  the tunneling probability
$P_t$ through the barrier in channel $U$ for the case $\Delta > 0$. The textbook expression for the transmission probability through a barrier is 
\begin{eqnarray}
P_t &=& \exp \left(-2\frac{\sqrt{2\mu}}{\hbar}\int_{Tp1}^{Tp2} \sqrt{V-E_{coll}} \, dr \right)\,,
\label{eq:tunnel_gen}
\end{eqnarray}
where $T_{p1}$ and $T_{p1}$ are the classical turning  points.
As in section \hyperref[sec:HarmBarrierModel]{E} of the Appendix 
we make the approximations that $\Omega$ is independent of $r$, that $V_e =  -0.5 \, k_{H} \, (r-r_s)^2 - \hbar \Delta$ is simply harmonic and that  $V_g = 0$. For $\Delta > 0$, the adiabatic potential $V_U$ then approximately has the shape of a Lorentzian,
\begin{equation}
V_U(r)\approx  V_0 \frac{( \Gamma/2)^2}{( \Gamma/2)^2+ (r-r_s) ^2}\,,
\end{equation}
with height $V_0 = 0.5 (-\Delta + \sqrt{ \Omega^2 + \Delta^2)} \hbar$ which for $\Omega \ll \Delta$ goes over into the well-known expression for the light shift,  $V_0 = \Omega^2\hbar/(4\Delta)$.

The width of the barrier is given by $\ \Gamma = \sqrt{2 \Omega^2 \, \hbar / (V_0 \, k_{H}) } $
which for the same limit, $\Omega \ll \Delta$, goes over into $ \Gamma = \sqrt{8 \hbar \Delta / k_{H}} $. The classical turning points are $-T_{p1}= T_{p2} = \frac{ \Gamma}{2}\sqrt{\frac{1}{\beta}-1} + r_s$, using $\beta=E_{coll}/V_0$. The integral of Eq.$\:$(\ref{eq:tunnel_gen}) can be analytically solved, yielding
\begin{equation}
P_t = \exp \left( - \sqrt{ \frac{2\mu V_0}{\beta}} \,  { \Gamma (1- \beta )\over \hbar } \, C(\beta)  \right)\,,
\end{equation}
where
\begin{eqnarray}
C(\beta)  &=&  \int_{-1}^1 \sqrt{  (1-y^2) \over \gamma \, y^2 + 1 } \,  dy \nonumber \\
&=& -{ 2 \over \gamma} \left[ E(-\gamma) - (\gamma + 1) \, K(-\gamma) \right]\,.
\end{eqnarray}
Here, $\gamma = (1-\beta)/ \beta$, and the functions $K$ and $E$ are the complete elliptic integrals for the first and second kind, respectively. $C(\beta)$ can be approximated by the simple expression  $ C(\beta) \approx  0.226 \ln(\beta) + \pi/2 $ in the relevant range  $0.01 \le \beta \le 0.99$.

\subsection{Harmonic barrier model}
\label{sec:HarmBarrierModel}
We consider here the special case where the potential barrier $V_e$ [see e.g. Fig.~\ref{fig3}(a) and (c)] is purely harmonic and radially symmetric, i.e. $V_e = -0.5  k_H (r-r_s)^2 - \Delta \hbar $, and the coupling $\Omega$ between ground and excited state does not depend on $r$. We ignore the  $1/r^4$ dependence of the polarization potential of the ground state and set $V_g = 0$. The radial Schr\"odinger equation for $s$-waves then reads
\begin{equation}
-{\hbar^2 \over 2\mu } {d^2 \over dr^2 } \Psi +
\left( \begin{array}{cc}
0  &  \hbar\Omega /2 \\
\hbar\Omega /2	\, \, \, \,   &   V_e(r)
\end{array}\right)\,  \Psi= E_{coll} \Psi\,.
\label{eq:Vapprox}
\end{equation}
Similar as for a harmonic oscillator, "Hooke's constant" $k_H$ introduces an energy scale $E_s = 0.5 \hbar \sqrt{k_H / \mu}$, and a length scale $l_s = \hbar \sqrt{1/ (2 \mu E_s)}$. In units of these two scales  the Schr\"odinger equation becomes
\begin{equation}
- {d^2 \over d\tilde r^2 } \Psi +
\left( \begin{array}{cc}
0  &  \tilde \Omega / 2 \\
\tilde \Omega /2	\, \, \, \,   &   -\tilde r^2- \tilde \Delta
\end{array}\right)\,  \Psi= \tilde E_{coll} \Psi\,,
\label{eq:Vreduz}
\end{equation}
 where $\tilde r = (r - r_s)/l_s $, $\tilde \Omega = \hbar \Omega / E_s $, $\tilde \Delta = \hbar \Delta / E_s $, and $\tilde E_{coll} = E_{coll} / E_s$. Thus, the solution of the problem, along with the transmissivity and reflectivity of the barrier, and the non-adiabaticity of the crossing, only depend on the three dimensionless parameters $ \tilde \Omega, \tilde \Delta, \tilde E_{coll}$.

\end{document}